# Using SOA with Web Services for effective data integration of Enterprise Pharmaceutical Information Systems

Quist-Aphetsi Kester, MIEEE
Faculty of Informatics, Ghana Technology University College
Kquist-aphetsi@gtuc.edu.gh/ kquist@ieee.org

Ajibade Ibrahim Kayode
Graduate School Coventry University, U K / Ghana Technology University College
kayvaldo@yahoo.co.uk

*Abstract*— *Medicines exist for combating illnesses and drug stores provide an outlet for consumers to gain access to these drugs. However, due to various factors such as cost of production, accessibility and legal issues as well as distribution factors, availability of these drugs at any location cannot be guaranteed at all times. There is need for individuals and organizations such as hospitals to be able to locate required drugs within given geographical vicinity. This is of immense importance especially during emergencies, travel, and in cases where the drugs are uncommon.*
*This research work is aimed at solving this problem by designing a system that integrates all drugstores. The integration of the drug stores will be based on SOA concepts with web services via a central service bus. The database systems of the drug stores will be integrated via a service bus such that drugs can easily be searched for and the results will be displayed based on its availability. Drugs can easily be searched for within geographically distributed pharmaceutical databases as well as consumption of drugs with relation to geographical locations can easily be monitored and tracked. This will make it easy for health institutions to research on drug consumption patterns across geographical areas and also control their usage.*

*Keywords— SOA, web services, integration, drug discovery, pharmaceutical information systems*

## I. INTRODUCTION

To achieve the goal of the global national e-health strategy that intends to provide an interoperable, standardized and secure platform for all involved partners in supporting healthcare services, one needs to focus on interoperability and integration of all distributed enterprise e-health systems.

Service Oriented Architecture (SOA) has radically changed the application integration landscape. SOA can be considered as a business-centric approach for enabling integration. Visibility, interaction and effect are the key concepts in any SOA implementation [1]. Visibility refers to the capacity of those with a need to see those with a capacity to service the needs. This is accomplished by publishing the service in a registry.

The fundamental driver for any integration effort is the need to share information within the enterprise and with business partners. Progressively it is becoming evident that enterprises that have most effective information sharing are the ones that remain competitive and successful. Due to the large heterogeneity of the IT landscape in most large enterprises, enabling application integration is not an easy task. Application software integration, as a field, has received considerable attention over the past two decades as a result of expanding business requirements (a product of functional silos built around disparate and disconnected systems) and technological advances aimed at eliminating this challenge, once and for all. [2]

Enterprise Application Integration (EAI) is the process of linking different applications together within a single organization or across organization boundaries in order to simplify and automate business processes to the greatest extent possible, while at the same time avoiding having to make changes to the existing applications or data structures. In the words of the Gartner Group, EAI is the unlimited sharing of data and business processes among any connected application or data sources in the enterprise [3].

As corporate dependence on technology has grown more complex and far reaching, the need for a method of integrating disparate applications into a unified set of business processes has emerged as a priority. Users and business managers are demanding that seamless bridges be built to join them. In effect, they are demanding that ways be found to bind these applications into a single, unified enterprise application. The development of Enterprise Application Integration (EAI), which allows many of the stovepipe applications that exist today to share both processes and data, allows us to finally answer this demand. [4]

In today's enterprise infrastructure, system and application integration is a critical concern. Enterprise Application Integration has existed since the early 2000s, but the central problem that it attempts to solve is much older. EAI is an approach, or more accurately, a general category of approaches, to providing interoperability between the multiple disparate systems that make up a typical enterprise infrastructure. Enterprise architectures, by their nature, tend to consist of many systems and applications, which provide the various services the company relies upon to conduct their day to day business. A single organization might use separate systems, either developed in-house or licensed from a third party vendor, to manage their supply chain, customer relationships, employee information, and business logic. [5]

Service Oriented architecture (SOA) has gained popularity in recent years due to its enabling functionality or services to upgrade and extend existing software applications. SOA is an architectural approach to build and deploy software applications that is interoperable by design. SOA has grown as companies endeavor to leverage their existing client base and to integrate their acquired software with their clients' existing ERP system and also it makes software connectivity capabilities very easy. Unlike EAI no middleware is needed as adoptions of standards enable services to interact directly.

It also enhances reusability capacity of software, resulting longer life of existing assets. A successful SOA implementation makes it easier to customize and upgrade existing applications thereby reducing total cost of ownership. [6]

EAI products prove to be expensive, consume considerable time and effort and are subject to high project failure rates. Additionally, because these custom applications are proprietary, many of the projects result in additional difficulties. Importantly, modifications to such applications require developing almost the entire system from scratch. Recent experience shows that a better answer is available by using Web services standards. The adoption of service oriented Architecture and web services provide a rapid solution to solving this problems faced by organizations [7] [8] [9].

The paper has the following structure: section II consist of related works, section III gives information on the methodology, section IV discusses the approach used for the integration, V talks about implementation as well as results and section concluded the paper.

## II. RELATED WORKS

In this development paradigm, functionality is exposed as services thereby enabling service requesters and providers interact through messages. Services are built to be autonomous but can also be combined to form even larger services and applications. Service orientation provides guidelines and principles that govern the creation, implementation and management of services.

Primarily, services are implemented as Web Services (WS) which are defined by the W3C as "software systems designed to support interoperable machine-to-machine interaction over a network" [10].It has an interface described in a machine-processable format. Other systems interact with the Web service in a manner prescribed by its description using SOAP-messages, typically conveyed using HTTP with an XML serialization in conjunction with other Web-related standards [10].Extensible Markup Language (XML) has emerged as a powerful self-describing language to enable businesses to share information and conduct transactions on the Internet. The emergence of XML as a standard, to a large extent, has driven the evolution of application integration technologies. [1]

Inherently, a service is a software component that contains a collection of related software functionalities reusable for different purposes [11].It delivers such operations as data storage, data processing, mathematical and scientific computations, and networking. It is governed by a producer-consumer model in which a service is delivered by a service provider known as the producer which owns the facilities for hosting, running, and maintaining the service, and the client known as the consumer which connects and uses service functionalities.

Much of the research regarding SOA tackles more granular technical issues of development and implementation of Web services, which may be a result of the aforesaid misconceptions [12]. Few papers e.g., [13],[14], deal with the much larger problem of defining what SOA means to the organization and how this definition should then provide the guidance for the development of components to meet business information needs[15]. The IT adoption literature targeting a methodology for development states that there are five categories of factors influencing the decision to adopt SOA (i.e., environmental, organizational, individual, technology, and task characteristics [16]. These same factors should be addressed by the methodology for implementing SOA projects [17]. We now discuss two SOA methodologies that attempt to embody some or all of these factors.

Teti (2006), an industry analyst, provides a methodology, which entails creating a vision, construction, and execution. He suggests that this model is applicable to many projects, but specifically addresses SOA. The vision creation is driven by a number of inter- and intra-organizational issues that define tasks important to the individuals and the firm (i.e., the constituency); the construction addresses the technology required to accomplish the tasks; and execution seeks to ensure that SOA will facilitate information exchange in the environment.

Bell (2008) provides a SOA methodology that takes a more technical approach. It professes that all software can be considered as services that are designed based on the informational tasks of the organization, configured for transmission in the working environments, constructed with available technologies, and deployed for use by individuals. The methodology represents a conceptual structure that brings together distributed services based on the functionality [18].

## III. METHODOLOGY

Providing services through the Web is rapidly becoming the emerging trend. Enterprises are recognizing that it is important for them to provide more of their services, such as customer support and product catalogs, through the Web. Enterprises have come to see that having such services available both in a traditional manner and over the Web enhances their business. The technology scenario is evolving at a breathtaking pace, and EAI is now increasingly being driven by Web-driven requirements and technologies. Web-driven application integration, by making data and services more easily and widely accessible enhances efficiency in business processes.

The paper is aimed at integrating distributed pharmaceutical information systems based on SOA concepts with web services via a central service bus. The database systems of the drug stores will be integrated via a service bus such that drugs can easily be searched for and the results will be displayed based on its availability and closeness to the search point. This will be done by designing architectures to the problem of discovering drug availability. Service-oriented architecture is the overarching paradigm employed and web services technology is used to implement the service orientation principles. The enterprise service bus provides a means of orchestrating and managing services throughout the systems lifecycle.

IV. **THE INTEGRATION APPROACH**

A multi-tier architecture is employed in devising the solution. The architecture consists of the database layer, business logic layer, middleware/ service bus, and presentation layer. The database and business logic layer reside at the service providers' end. The database contains information relating to drugs, some of which would be made available to consumer while the business logic layer is where the business rules that determine how data is created, stored and accessed are implemented. The database technologies and business logic implementations vary across service providers. Access to the database is via the business logic layer. As a result of the service-oriented architecture being implemented, external access to these databases is provided via services. The services are implemented using web service technology. The choice of web services technology is due to its support for open technologies and protocols such as HTTP (hypertext transfer protocol), XML (extensible markup language), and, WSDL (web service definition language) among others. These open technologies provide a desired level of interoperability and easy means of access. The web services provide functionality to access the databases while implementing the business logic at the backend. The individual services are registered in a repository from where they can be accessed. While consumers can access services directly in repositories, in this case, access will be through a middleware. An enterprise service bus is the middleware between service consumers and the services. Since the consumers need to access various services to get comprehensive information on drugs, the ESB performs the task of determining what service to access and the order they are accessed. In other words, the ESB performs service orchestration. Service consumers cannot access the ESB directly so a further interface is required. Service consumers request services today via an array of devices ranging from desktop computers and laptops to mobile devices such as smartphones and tablets. A universal mode of access is required for all these devices hence the choice of a web interface. Most modern devices contain a web browser that can be used to access web content irrespective of operating platforms. As a result, a web server is required to provide a presentation layer in form of web pages to the consumers via which they can access the services and view results.

This choice of architecture is effective because it provides access to a wide range of consumers and devices as a result of the ubiquitous technologies employed in the presentation layer while also providing a high level of separation of concern. However, the consumers are loosely coupled to the individual services. This provides flexibility because the service providers can switch database technologies or business logic implementation without affecting the consumer as long as the exposed services conform to a predefined contract that the consumer is accustomed to. Implementation of services by providers is also relatively cheap as services can be built on existing infrastructures without restructuring the whole system.

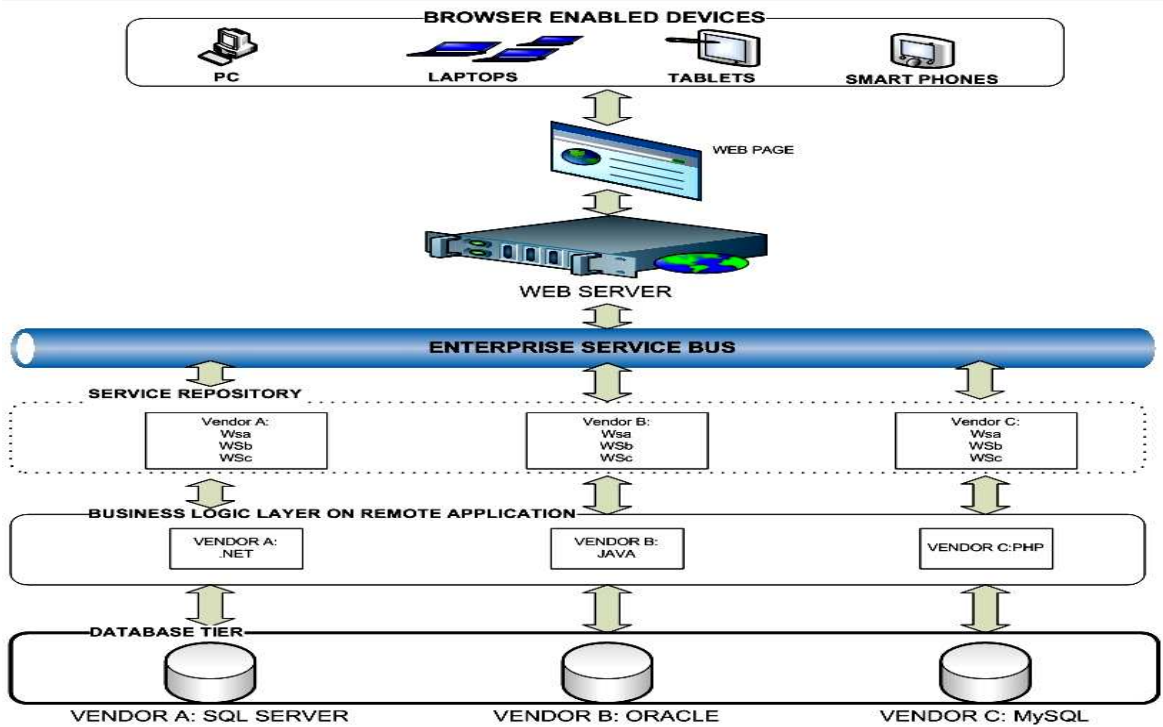

Fig. 1 Integration the architecture layers and infrastructure

From fig 1, a central body defines the services to be exposed by drug vendors (represents drug stores and health institutions). These could include services to determine:
- That a drug is available in stock.
- The quantity available.

- The price of the drug.
- Vendor information such as address in case the client wants to patronize the store.

Consumers seeking to use this service are provided with a front end web page through which they can search for information regarding drugs. The search criterion is primarily the drug name but this can be varied. The search input is posted back to the web server on submission and then the input is transferred to the enterprise service bus via the appropriate adapter. The ESB has a collection of approved web services exposed by the various vendors in its registry. The ESB passes the input parameters to the appropriate services in the registry. These services in turn transfer the input parameter to their remote applications on which their business logic resides. The applications query the databases using the supplied parameter and the results are passed back to the ESB. The results are accumulated and transferred back to the web server where they are formatted and displayed to the end user in a web page.

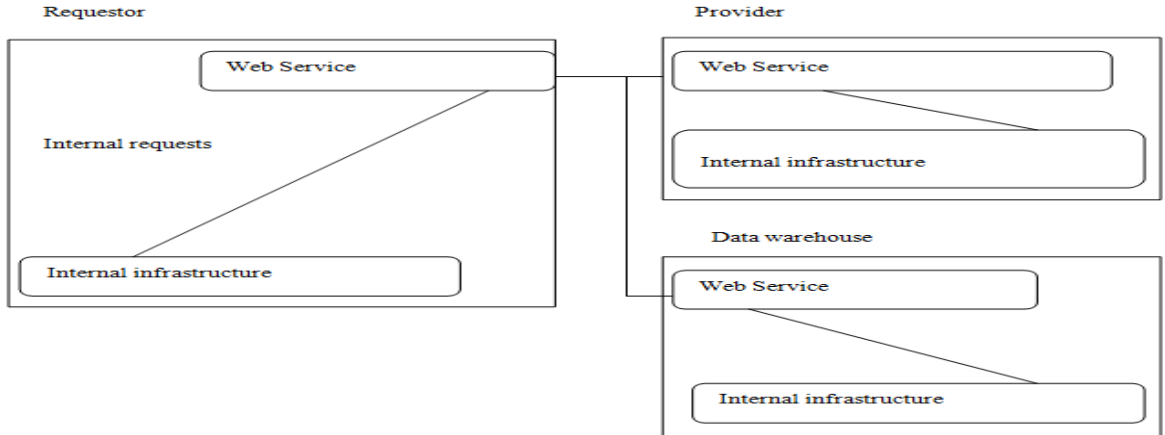

Fig. 2  Web enabled systems for provider and requestor.

In the figure above, languages and protocols were standardized to eliminate the need for many different middleware infrastructures and the interactions were based on protocols redesigned and the internal functionality settings were made available as a service. Service-oriented architecture has standardization as a key policy in implementing web services for easy integration of multiple incompatible applications [19]. Web services provide an entry point for accessing local services and with homogeneous components that reduces the difficulties of integration. Web services were exposed through the interface. Homogeneous components were built to reduce the difficulties of integration. Service descriptions were made richer and more detailed, covering aspects beyond the service interface.

The architecture of the solution proposed in this paper is based on the Open Group SOA Reference Architecture as seen in fig 3. The SOA reference architecture provides a baseline that shows the basic layers involved in a typical SOA solution. This diagram shows the different layers of the reference model and how they fit together to provide a standard service oriented solution. The Open Group SOA Ontology provides a taxonomy and ontology for SOA. [20]

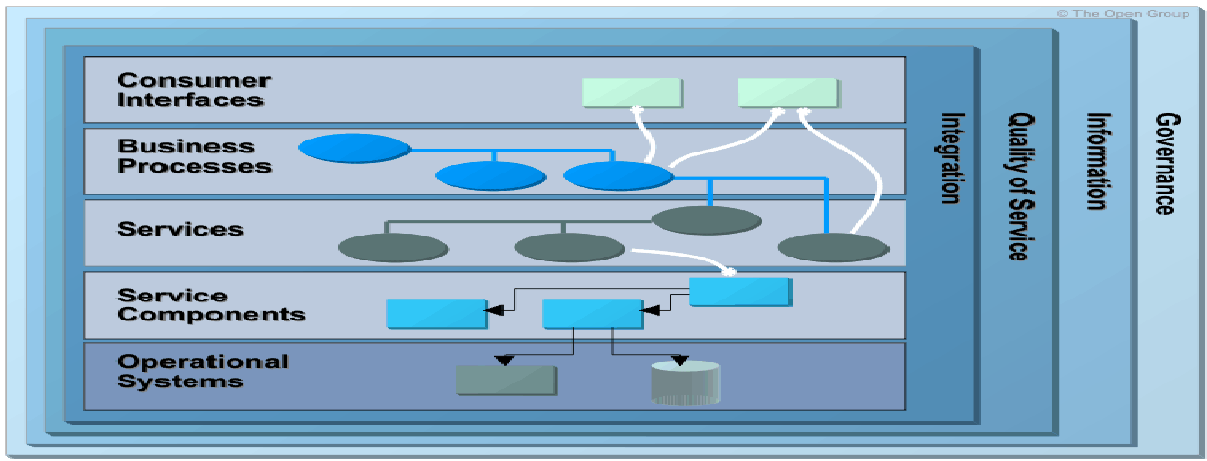

Fig. 3  The Open Group SOA Reference Architecture.

Fig. 4 illustrates how various disparate data sources within pharmaceutical information systems are connected via a service bus. Requests can be made to check the availability of specific drug in shops. Institutions and individuals can perform searches using Smartphone and other devices.

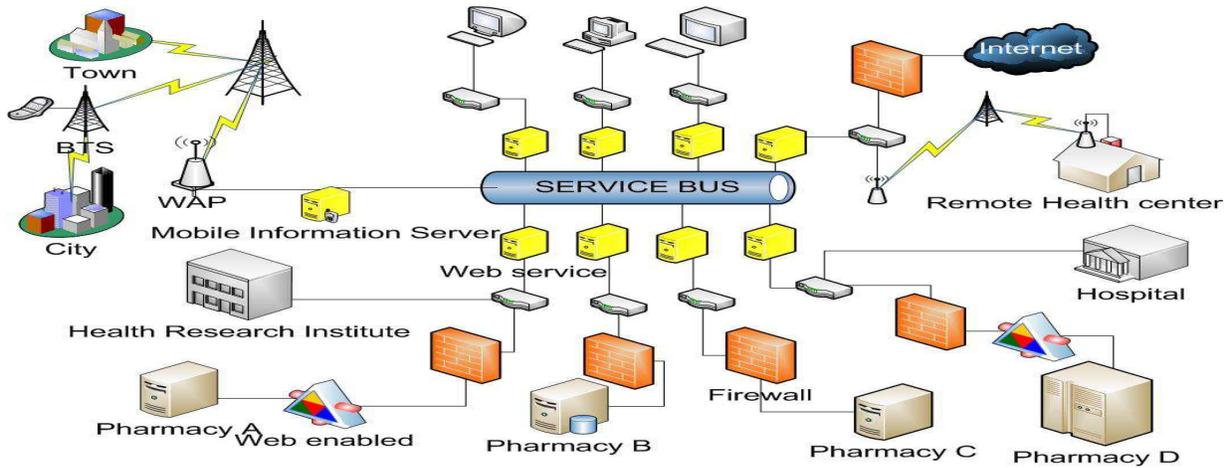

Fig. 4  Integrated Systems via a Service Bus.

V. **IMPLEMENTATION**

The aim of this section is to demonstrate using Windows Communication Foundation and the Entity Framework, how a service can be created to return information resident in a database about a drug and how that information can be constructed to conform to a predefined specification.

The following information is required to be retrieved based on the name of the drug that is passed as a parameter:
- The price of a drug
- A description of the drug
- The name of the vendor

This information is appended with the name of the drug when the results are returned. Other relevant information such as the address of the vendor's outlet and available substitutes can be further requested when the consumer decides. The specification is represented in form of an XML schema file shown below;

```
DrugInfoSpecs.xsd
    <?xml version="1.0" encoding="utf-8"?>
    <xs:schema attributeFormDefault="unqualified" elementFormDefault="qualified" xmlns:xs="http://www.w3.org/2001/XMLSchema">
        <xs:element name="Drug">
            <xs:complexType>
                <xs:sequence>
                    <xs:element name="name" />
                    <xs:element name="Price" />
                    <xs:element name="Description" />
                    <xs:element name="VendorName" />
                </xs:sequence>
            </xs:complexType>
        </xs:element>
    </xs:schema>
```

Fig. 5  schema files/ result specification.

This shows one of the possible ways that a solution can be implemented to return the information that matches the schema above. This solution assumes that the provider's existing solutions are based on the .Net platform which includes an instance of SQL Server and a Microsoft operating system. The following steps were ensured to implement the service:

STEP 1: A WCF service was created and used to retrieve the required information. The functionality to be exposed was specified as GetDrugInfo(string drugName). The operation contract attribute was appended to it to specify that it was an operation to be exposed to external consumption via a specified endpoint or set of endpoints. In order for this service to be accessed via a URI (uniform resource indicator), the WebGet attribute was appended to the exposed service to enable it to be accessed. Using Representational State Transfer also ensured that the results were returned in POX (plain old XML). The UriTemplate property of the WebGet attribute was used to specify the additional information that was appended to the basic address in order to access the service.

```csharp
namespace ProjectServiceLibrary
{
    // NOTE: You can use the "Rename" command on the "Refactor"
    [ServiceContract]
    public interface IDrugInfoService
    {
        [OperationContract]
        [WebGet (UriTemplate="GetDrugInfo/{drugname}")]
        DrugInfo GetDrugInfo(String drugName);

        // TODO: Add your service operations here
    }
}
```

Fig. 6  The exposed service and corresponding attributes.

The DrugInfo class was created to match the predefined schema. This was necessary to ease information mapping between the database and the specification. The class was appended with the DataContract attribute which specifies that it is a data type to be transported. The properties were marked with the DataMember attribute which specifies how they are serialized in XML.

```csharp
[DataContract]
public class DrugInfo
{
    string drugName;
    string drugDescription;
    string vendorName;
    decimal drugPrice;

    [DataMember]
    public string DrugDescription
    {
        get { return drugDescription; }
        set { drugDescription = value; }
    }

    [DataMember]
    public string VendorName
    {
        get { return vendorName; }
        set { vendorName = value; }
    }

    [DataMember]
    public string DrugName
    {
        get { return drugName; }
        set { drugName = value; }
    }

    [DataMember]
    public decimal DrugPrice
    {
        get { return drugPrice; }
        set { drugPrice = value; }
    }
}
```

Fig. 7  The druginfo class created to match the schema specifications.

STEP 2: The aim here was to provide the basic data retrieval functionality. An object relational mapper was used to generate classes that can be worked with in order to retrieve information from the database. In this situation, the Entity Framework was used. Since the database already existed, the "database first" model of the entity framework was used to generate a data model and the corresponding classes.

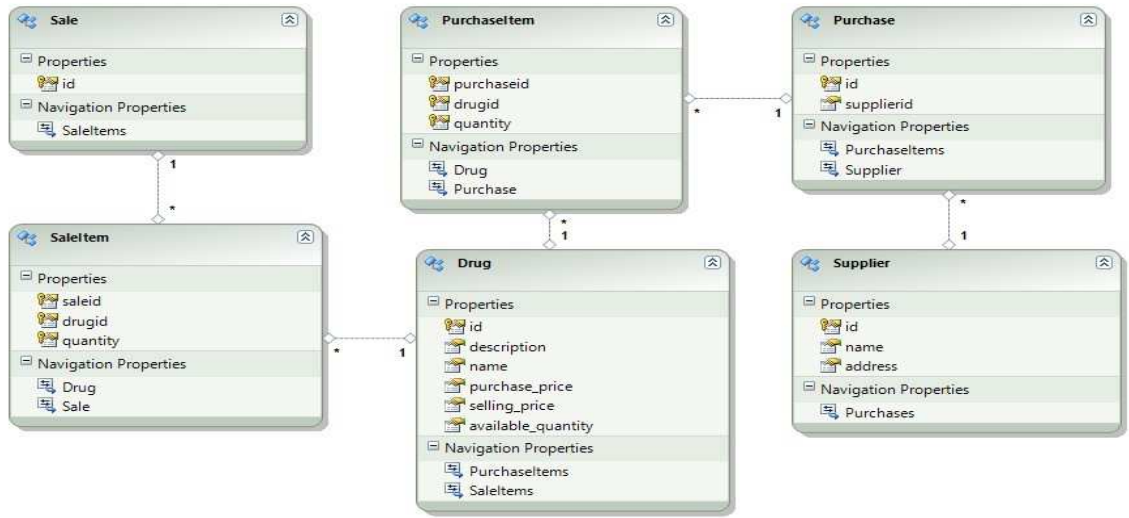

Fig. 8 The data model generated from the existing database.

STEP 3: Queries were created to retrieve the specified information based of the name of the drug passed as parameter. The LINQ to ENTITIES query model was used to create these queries. The results of the queries were used to create an instance of the class that corresponds to the schema specification.

```
//Implementation of Service Method
public DrugInfo GetDrugInfo(string drugName)
{
    var context = new Zoch_PharmacyEntities1();
    var drug = from c in context.Drugs
               where (c.name.Equals(drugName))
               select c;
    return new DrugInfo { DrugName = drug.First().name,
        DrugDescription = drug.First().description,
        DrugPrice = drug.First().selling_price,
        VendorName = "Zoch Pharmacy" };
}
```

Fig. 9 The database access code using LINQ to entities.

The service was hosted on a web server and a URL was provided through which the web service can be accessed. A parameter representing the name of a drug was appended to the URL in order to retrieve the relevant information pertaining to that drug. The name used in this example was "Blopen Gel". Below is the corresponding response received formatted in XML and the result corresponds with the initial schema specified.

```
localhost:8732/drugservice.svc/getdruginfo/blopen%20gel

This XML file does not appear to have any style information associated with it. The document tree is shown below.

<Drug xmlns="http://schemas.datacontract.org/2004/07/ProjectServiceLibrary" xmlns:i="http://www.w3.org/2001/XMLSchema-instance">
  <Description>Deep penetrating gel for aching joints and muscles</Description>
  <Name>Blopen Gel</Name>
  <Price>5.0000</Price>
  <VendorName>Zoch Pharmacy</VendorName>
</Drug>
```

Fig. 10 The druginfo class created to match the schema specifications.

The XML results were collated from different sources and then a web page generated with this information which was presented to the consumer. It should be noted that no currency symbol is attached to the price in the diagram above in order to facilitate conversions between different currencies. The symbol can be added in the final presentation to the customer.

VI. CONCLUSION

With the proposed system, individuals as well as health institutions can now search for drugs from databases that have been web enabled and have their services registered within the service repository which is discoverable via the service bus. Using SOA with web services makes it easy for heterogeneous database platforms to be integrated and interoperate. Services created can be reused in multiple and also new services and applications can be created quickly and easily used with a combination of new and old services.